# Measurement of the polarization for soft x-ray magnetic circular dichroism at the BSRF beamline 4B7B[*]

GUO Zhi-Ying(郭志英)[1,2] XI Shi-Bo(席识博)[1] ZHU Jing-Tao(朱京涛)[2] ZHAO YI-Dong(赵屹东)[1]  ZHENG Lei(郑雷)[1]

HONG Cai-Hao(洪才浩)[1]  TANG Kun(唐坤)[1]  YANG Dong-Liang(杨栋亮)[1]  CUI Ming-Qi(崔明启)[1;1)]

1 Institute of High Energy Physics, Chinese Academy of Sciences, Beijing 100049, China

2 Institute of Precision Optical Engineering, Physics Department, Tongji University, Shanghai 200092, China

**Abstract**：Three ultra-short-period W/$B_4$C multilayers (1.244nm, 1.235nm and 1.034nm) have been fabricated and used for polarization measurement at the 4B7B Beamline of Beijing Synchrotron Radiation Facility (BSRF). By rotating analyzer ellipsometry method, the linear polarization degree of light emerging from this beamline has been measured and the circular polarization evaluated for 700eV-860eV. The first soft x–ray magnetic circular dichroism measurements are carried out at BSRF by positioning the beamline aperture out of the plane of the electron storage ring.

## 1 Introduction

Soft X-ray synchrotron radiation has excellent polarization characteristic. Many experiments, such as X-ray magnetic circular dichroism (XMCD) [1] and magnetic linear dichroism (XMLD) [2], benefit from high degree linear or circular polarized light. XMCD is the asymmetric absorption of left- and right-handed circularly polarized X-rays and can be used to measure element-specific spin and oxidation states and magnetic moments [3]. The dichroic signal is proportional to the circular polarization degree which has to be perfectly known in order to interpret quantitatively the dichroic signals. In addition, a depolarization effect from reflection and diffraction by the beamline optics may influence and even degenerate the degree of polarization, so it is necessary to measure the polarization states of the light emerging from monochromator.

In order to realize a full polarization measurement, a phase retarder and analyzer are needed. In the vacuum-ultraviolet region, reflection mirrors are usually used as retarders and analyzers [4, 5]. In

\_\_\_\_\_\_\_\_\_\_\_\_\_\_\_\_\_\_\_\_\_\_\_\_\_\_\_\_

\* Supported by National Natural Science Foundation of China(11075176,10435050)

1)E-mail: cuimq@ihep.ac.cn

2)E-mail: zyguo@ihep.ac.cn

soft X-ray region a variety of transmission and reflection multilayers have been developed and utilized for polarimeter, such as Mo/Si for 50eV~100eV [6-9], Ru/Si [10] for~97eV, Ru/C [11] for~140-190eV, Cr/C [12] for~265eV, Cr/Sc [13] for ~400eV. However, for energy range 600eV-800eV, it is difficult to make polarization analysis due to the lack of suitable ultra-short-period multilayers. Recently short period W/B4C has been fabricated and used for polarization analysis [14, 15].

In our present study, a W/B4C multilayer has been developed as an analyzer and used in a compact polarimeter [16]. This polarimeter can be operated under double-reflections modes for full polarization measurement. However, due to the difficulty of alignment and the low reflectivity by two multilayers, the method of rotating only one analyzer has been employed if we only care about linear polarization degree and evaluates the circular polarization for MCD experiment. By rotating analyzer ellipsometry (RAE) method, the linear polarization degree of light has been measured. Linear polarization degree changing with slit width (entrance slit width from 100 to 200 microns) has been observed. Then, the dependence of polarization state on the vertical observation angle which varies has also been studied. Besides, we have evaluated the circular polarization at the L edge of Co film on the assumption $V=1$ ($V$ is the degree of polarization). At the end of this paper the first soft XMCD measurements at BSRF are reported.

## 2 Analytical method

Any state of light can be described using the normalized Stokes parameters which can be expressed as:

$$\begin{pmatrix} S_0 \\ S_1 \\ S_2 \\ S_3 \end{pmatrix} = \begin{pmatrix} 1 \\ V\cos 2\varepsilon \cdot \cos 2\delta \\ V\cos 2\varepsilon \cdot \sin 2\delta \\ V\sin 2\varepsilon \end{pmatrix}. \qquad (1)$$

where $\varepsilon$ is the ellipticity angle, $\delta$ is the azimuthal angle of the major axis of the polarization ellipse and $V$ is the degree of polarization. The arrangement of this method is illustrated in Fig. 1.

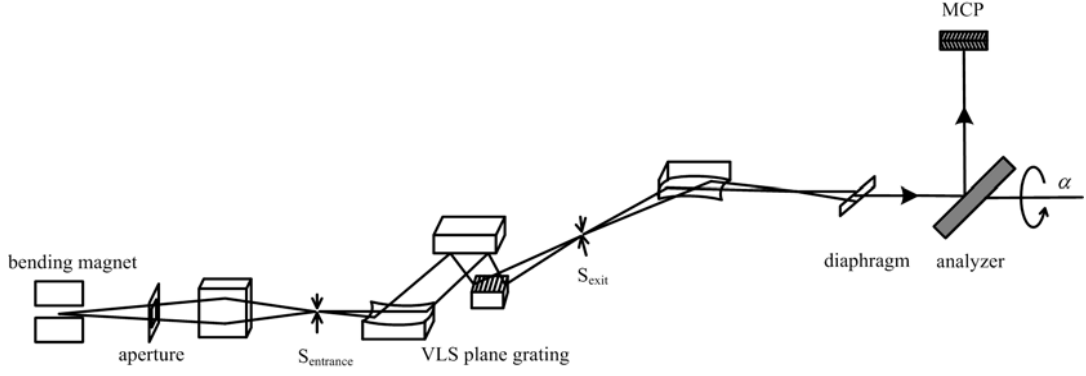

Fig. 1. Side view of the setup of the measurement

By changing the azimuth $\alpha$, the reflection intensity $I(\alpha)$ is recorded, which can be written as:

$$I(\alpha) = (\kappa^2 + 1)S_0 + (\kappa^2 - 1)(S_1 \cos 2\alpha + S_2 \sin 2\alpha). \tag{2}$$

and $\kappa$ is a ratio of amplitude reflectance for the s to p component, $r_s/r_p$. By fitting Eq. (2) by means of a least-squares method to the measured data, we can finally determine the parameters $\{S_0, S_1, S_2, \kappa\}$. The degree of linear polarization $P_L$ and the azimuthal angle of the major axis of the polarization ellipse $\delta$ are defined using the normalized Stokes parameters as follows:

$$P_L = \frac{\sqrt{S_1^2 + S_2^2}}{S_0} = V \cos 2\varepsilon, \qquad P_C = \frac{|S_3|}{S_0} = V \sin 2\varepsilon. \tag{3}$$

$$\delta = \frac{1}{2} \arctan \frac{S_2}{S_1}. \tag{4}$$

## 3 Experiment

Polarization measurements were carried out at the 4B7B Beamline of Beijing Synchrotron Radiation Facility (BSRF). 4B7B is a soft X-ray beamline with a bending magnet, which uses a variable-included-angle Monk-Gillieson mounting monochromator with a varied-line-spacing plane grating covering the energy range of 50eV~1600eV. The polarimeter was installed at the back of the end station.

Three short period W/B4C (1.244nm, 1.235nm and 1.034nm) multilayers were designed (for energy 708eV, 713eV, and 851eV) at the quasi-Brewster angle and they were deposited on Si wafers using a high vacuum direct current (DC) magnetron sputtering system. The reflectivity of the multilayers was measured with a reflectometer before they were used. Fig. 2 shows the measured peak reflectivity of about 2.2% for the multilayer of 1.235nm. The experimental result was simulated using the IMD 4.1.1 program. A thickness ratio 0.4 W/B4C structures with 200 periods and 0.3nm roughness were used for the calculations. The incident angle was 46.57deg and the polarization $P$=0.8 was assumed. The fit does not include energy resolution and divergence of the incident light. The ratio $\kappa$ which value is 282 can be calculated using the above fitting parameters.

2% reflectance was measured for multilayer with a period of 1.244nm. For the multilayer (1.034nm) used for 851eV, the reflectance was only 0.72% due to interface roughness and diffusion. Then the polarization state was measured with the polarimeter using the RAE method for 709eV. The W/B4C multilayer (1.235nm) was used as analyzer and a chevron arrangement of two micro-channel plates (MCPs) was used as detector. The incidence angle of the multilayer was about 45 deg, which was referred to a vertical position determined by means of auto-collimation. The azimuth was rotated manually in a range of $-90° < \alpha < 270°$, at the same time the reflected intensity was recorded with every interval of 3 degrees. Fitting the experimental data, we determined the Stokes parameters and $\kappa$: $\{S_0, S_1, S_2, \kappa\} = \{0.0378, 0.0275, -5.6\times10^{-6}, 257.05\}$. This shows the linear polarization degree is $0.729 \pm 0.006$ at 709eV. Fig. 3 shows the fitting sine curve and measured data.

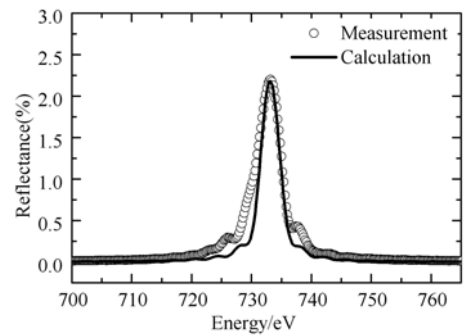

Fig. 2. Reflectance of a W/B4C multilayer as function of the photon energy. The symbol: experimental data. The line: calculation curve.

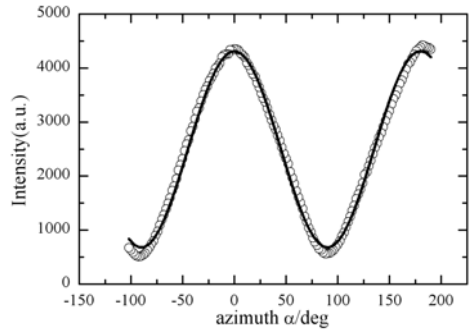

Fig. 3. The reflected intensity was recorded with the azimuth angle in a range of $-90° < \alpha < 270°$. The hollow circles show the measured data and the line shows the result of fitting.

It is well known that SR from a bending magnet is highly linearly polarized in the plane of the orbit and elliptically polarized away from the plane. So the dependence of the $P_L$ with the slits of the beamline has been studied. The entrance slit width $S_{entrance}$ and exit slit width $S_{exit}$ maintain the ratio: $S_{entrance} / S_{exit}=2$. The $P_C$ was calculated on the assumption $V=1$. As showed in Fig. 4, the $P_C$ increases with the slit width. The $S_{entrance}$ was set to be 180μm for XMCD experiments in order to obtain light with almost 70% circular polarization. The dependence of the polarization state on the vertical observation angle has also been studied. The light in different orbit plane has been selected by moving the 3×1mm rectangular diaphragm (3mm width in horizontal direction and 1mm in vertical direction) after the focus. In vertical direction 1mm diaphragm width corresponds to 0.185mrad. It can be observed from Fig. 5 that the linear polarization changes with the vertical position of the diaphragm. The $P_L$ is high in the middle of the orbit and the $P_C$ increases with the observation angle.

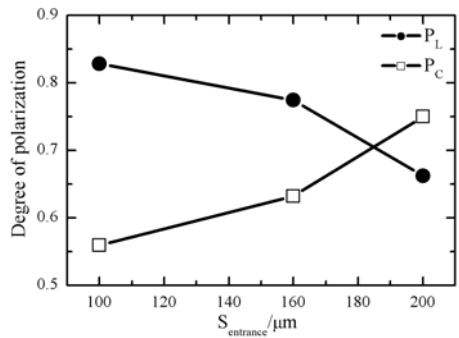

Fig. 4. The measured $P_L$ and the calculated $P_C$ as a function of the slits ($S_{entrance} / S_{exit}=2$).

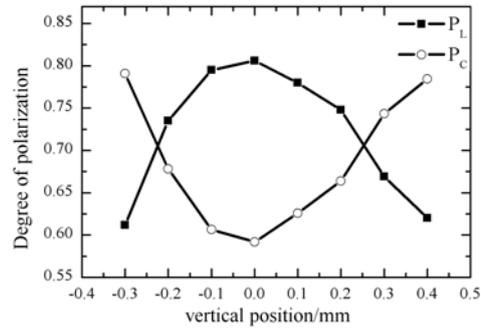

Fig. 5. Polarization dependence of the vertical position (or observation angle) of the diaphragm when the $S_{entrance}$ is 100 microns.

High degree circular polarized light is necessary for XMCD experiment. For bending magnet, the flux falls dramatically as observation angle increases. The photon flux and degree of polarization are correlated. In order to get the XMCD signal, the slit and the position of aperture were optimized. In order to ensure that $P_C$ >70% the slit $S_{entrance}$ was set to be 180 microns. Circular polarized light is obtained by positioning the beamline aperture out of the plane of the electron storage ring. The position of the aperture was optimized to make the dichroic signal maximum. We find that the optimum offset angle is ~0.6 mrad for XMCD experiments. At this angle the circular polarization rate is about 80%±5% and the relative intensity is still 50%.

In order to verify the circular polarized light obtained by the above method, the XMCD measurements were performed on 300nm Co film grown on Si by DC sputtering. For soft X-ray absorption spectroscopy (XAS), three methods are usually used: the total electron yield (TEY), the fluorescence yield (FY) and the transmission mode. TEY has a better intrinsic noise statistic compared with FY and lower sample requirements related to transmission mode, so the absorption signal was detected by an electrometer (Model 6517 A, Keithley, USA) limited to the pA range. To reduce the influence by external magnetic fields (side effect) and increase the signal-to-noise ratio, a ring electrode with bias voltages was placed in front of the sample. The sample surface was perpendicular to the incidence light. The right-handed circularly polarized light was obtained by moving the aperture down of the orbit plane for 7mm. Homemade permanent magnets were used to polarize the sample. The magnetic filed was about 4kOe and switched by manually reversing the magnets. XAS across the $L_{2,3}$ edges of Co was measured several times by fixing the polarization and inverting the applied magnetic field parallel or antiparallel to the direction of the incident light. Fig. 6 shows the XAS spectra and the respective difference XMCD signal. The signal is normalized by the TEY signal measured from a copper foil positioned under toroidal mirror. Applying the sum rules, we have determined the spin and orbit moment contributions for Co film: $m_{orb}$ ($\mu_B/atom$=0.02), $m_{spin}$ ($\mu_B/atom$=0.3). Due to surface oxidation and non-polarization, this value is lower than the theoretical result. In the near future, low temperature and electromagnetic system will be used, and it is expected that the result will be better.

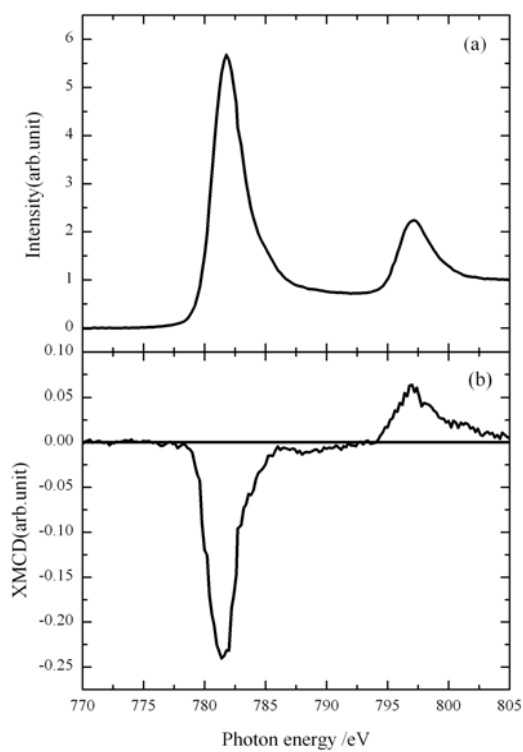

Fig. 6.　(a)XAS and (b) XMCD spectra of a 300nm Co film at normal x ray incidence.

## 4　Conclusions

Polarization measurements were carried out by RAE method with W/B4C multilayers at 4B7B Beamline. The $P_L$ is high in the middle of the orbit and the $P_C$ increases with the observation angle. Circular polarized light can be obtained by positioning the beamline aperture out of the plane of the electron storage ring. The optimum offset angle is ~0.6 mrad for XMCD experiments.

The soft XMCD was first carried out at 4B7B Beamline by TEY mode. A bias voltage electrode can improve the signal-to-noise ratio and eliminate the side effect. The permanent magnet limits the maximum magnetic field. The signal will be better in the future if the electromagnets can be used and combined with low temperature system.

*We would like to thank Dr. Wang Zhan-Shan and his group for supplying multilayer and Co film.*

*his work was supported by National Natural Science Foundation of China (11075176, 10435050).*

## References

1  Stohr J. J. Electron Spectrosc. Relat. Phenom., 1995, **75**: 253-272

2  Stohr J. Surf. Rev. Lett., 1998, **5**(6): 1297-1308

3  Funk T, Friedrich S, Yong A T et al. Rev. Sci. Instrum., 2004, **75**(3): 756

4  Gaupp A, Mast M. Rev. Sci. Instrum., 1989, **60**(7): 2215

5  Koide T, Shidara T, Yuri M et al. Appl. Phys. Lett., 1991, **58**: 2592

6  HU Wei-bing, Hatano T., Yamamoto M. et al. J. Synchrotron Rad. 1998, **5**: 732

7  Schafers F, Mertins H C, Gaupp A et al. Appl. Opt., 1999, **38**: 4047

8  WANG Hong-chang, ZHU Jing-tao, WANG Zhan-shan et al. Thin Solid Films, 2006, **515**: 2523

9  WANG Zhan-shan, WANG Hong-chang, ZHU Jing-tao et al. Appl. Phys. Lett., 2007, **90**: 031901

10  Kimura H, Yamamoto M, Yanagihara M et al. Rev. Sci. Instrum., 1992, **63**(1): 1379

11  Kortright J B, Rice M, Franck D K. Rev. Sci. Instrum., 1995, **66**(2): 1567

12  Fonzo S D, Jack W, Schafers F et al. Appl. Opt., 1994, **33**: 2624

13  Wagner U H. A Versatile Multilayer Polarimeter for the Soft X-Ray Region. In: Garrett R& Gentle I. Proc. of the 10th Int'l. Conf. on Synchrotron Radiation Instrumentation. Melbourne (Australia): American Institute of Physics, 2009. 781

14  WANG Zhan-shan, WANG Hong-chang, ZHU Jing-tao et al. Appl. Phys. Lett., 2007, **90**: 081910

15  MacDonald M A, Schaefers F, Pohl R et al. Rev. Sci. Instrum., 2008, **79**: 025108

16  CUI Ming-Qi. The soft x ray polarimeter and applications at BSRF. In: Garrett R& Gentle I. Proc. of the 10th Int'l. Conf. on Synchrotron Radiation Instrumentation. Melbourne (Australia): American Institute of Physics, 2009. 641